\documentclass{iopart}
\usepackage{amsthm}
\usepackage{amssymb}
\usepackage{graphicx}
\usepackage{multicol}

\newtheoremstyle{theorem}{\topsep}{\topsep}%
     {}
     {}
     {\bfseries}
     {}
     {0.3cm}
     {}
\theoremstyle{theorem}

\newtheoremstyle{example}{\topsep}{\topsep}%
     {}
     {}
     {\bfseries}
     {}
     {\newline}
     {\thmname{#1}\thmnumber{ #2}\thmnote{ #3}}
\theoremstyle{example}

\newcommand{\vect}{\vec}

\newcommand{\beq}{\begin{eqnarray}}
\newcommand{\eeq}{\end{eqnarray}}

\newcommand{\half}{\frac{1}{2}}

\newcommand{\lbr}{\left( }
\newcommand{\rbr}{\right) }

\newcommand{\lbrs}{\left[ }
\newcommand{\rbrs}{\right] }

\newcommand{\Dp}[2]{\frac{\partial #1}{\partial #2}}

\newcommand{\aeq}{\simeq}

\newcommand{\ignore}[1]{}

\newcommand{\cross}{\times}

\newcommand{\fancy}[1]{\mathcal{#1}}

\renewcommand{\Dp}[1]{\partial_{#1}}

\renewcommand{\vect}[1]{{\bi{#1}}}
\newcommand{\tens}[1]{\bi{#1}}
\renewcommand{\v}{\vect}
\newcommand{\vh}[1]{\hat{\vect{#1}}}
\renewcommand{\t}{\tens}

\newcommand{\T}{{\rm{T}}}

\newcommand{\comm}[2]{\lbrs #1, #2 \rbrs} 

\newcommand{\spc}{\hspace{0.3cm}}

\newcommand{\Nname}{{\rm{dist}}}
\newcommand{\N}[1]{\Nname(#1)}

\newcommand{\dv}[1]{\Delta \v{#1}}
\newcommand{\hv}[1]{\hat{\v{#1}}}

\newcommand{\ind}[1]{^{{\lbr #1 \rbr}}}
\newcommand{\cdotc}{}

\begin{document}

\newcommand{\micron}{$\rm{\mu m}$}
\newcommand{\sfrac}[2]{#1 / #2}
\newcommand{\vphi}{{\mbox{\boldmath $\mathbf{\phi}$\unboldmath}}}
\newcommand{\reg}{{${}^{\circledR}$}}
\newcommand{\tm}{{\texttrademark}}

\title[Defective beams in MEMS modelling]
{Defective beams in MEMS: a model of non-ideal rods using
a Cosserat approach for component level modelling}
\author{Tim Gould and Charles H-T Wang}
\address{Department of Physics, Lancaster University, Lancaster LA1 4YB}
\eads{\mailto{t.gould@lancaster.ac.uk}, \mailto{c.wang@lancaster.ac.uk}}

\begin{abstract}{
We present and derive a technique for the introduction of defects
into a beam model based on the Cosserat theory of rods. The
technique is designed for the derivation of component models
of non-ideal rods for use in MEMS devices. We also present a
worked through example of blob/nick defects (where the rod
has an area with an excess/lack of material) and a guide
for a model with random pits and blobs along the length of the
beam. Finally we present a component level model of a beam
with a defect and compare it to results from a Finite Element
Analysis simulation. We test the Cosserat model for two cases
without any defect and four with a defect. Results are in
good agreement
with a maximum $0.5\%$ difference for the ideal case and
under $1\%$ differences for all but one of the defective cases,
the exception being a $2\%$ error in an
extreme case for which the model is expected to break down.
Overall, the Cosserat model with and without defects provides
an accurate way of modelling long slender beams. In addition,
simulation times are greatly reduced through this approach
and further development for both component level models
as well as as FEA components is important for practical yet
accurate modelling of MEMS both for prediction and comparison.
}\end{abstract}

\maketitle

\section{Introduction}

Many MEMS devices rely on flexible rods of Si
(or other materials) to act as a `spring' for the device.
As such, accurate models of such rods have been developed
\cite{MEMS1, MEMS2, Wang:Cosserat2D, Wang:Cosserat3D}
to facilitate simulation of these devices.
Although these techniques can be quite sophisticated, so far work has
concentrated on perfect (prismatic) rods i.e.
those whose cross-sections do not change along the length and which
have uniform material properties. While this is quite adequate
for most common MEMS, such as the Tang Resonator where the
system is designed to eliminate any \emph{minor} defects caused by
the manufacturing process, some more advanced devices may be
quite susceptible to defects as may present designs if
miniaturisation is to take place.

Presently, most simulation of such devices would be carried
out using Finite Element Analysis
through a package such
as ANSYS\reg\footnote{ANSYS\reg is a registered trademark of ANSYS, Inc.}
or FELT \cite{Felt}. These simulations can be extremely time
consuming both at the design level and during the calculation. Once
defects are added it could be expected that the time required
for design and to a lesser extent computation will
increase dramatically.

Another approach to modelling is component level (or network) modelling \cite{Component1, Component2}.
Here each component of a MEMS such as beams, shuttles,
comb drives etc. are modelled individually and joined
together. The resulting system requires finding the solution
of a system of differential-algebraic equations (DAEs) \cite{DAE}
which is a relatively simple task. As such, component level modelling
is very efficient for the modelling of new devices.
Another aspect of component models is that they can
also be introduced into FEA calculations to speed
computation of certain `regions' or `components' of the system.
Traditionally, beam models such as those of Euler-Bernoulli
or Timoshenko \cite{OldBeams} are used in these circumstances.

In this paper we propose a technique to model rods which
builds on a recent model of beams based on Cosserat Theory
\cite{Antman} to introduce non-ideal properties such
as defects caused by the manufacturing process.
To demonstrate the viability of this method we provide
two fully worked through and tested examples being a rod with
a blob of excess material (or a `nick' which is essentially the
same problem) and a rod in which the manufacturing process has
left a random jitter i.e. a series of minute pits and blobs
with a known statistical distribution.
We also run a series of tests on the blob/nick model where
FEA simulations are used as a benchmark of the model
developed in this paper.

The technique is designed to be easily implementable in a
component level simulation, allowing it to be used in rapid
modelling of present and future MEMS devices. The design
of a `defective component' can be carried out using this
approach and installed into a preexisting model.

\section{Preliminaries}

In the modelling of flexible rods, the Cosserat approach
is a good method for both analytic and numerical treatments. Through
a kinematic assumption that the cross-sections of the rod
change their orientation and position but not their shape along
the length of the rod (physically reasonable in most cases)
it is possible to write a Lagrangian for
a rod as follows \cite{Antman} (henceforth we use bold-italics for
matrices and vectors with capitals for matrices and lowercase for
vectors)
\begin{equation}
\label{eqn:Potentials}
\fl
\eqalign{
\fancy{T}&=\int_0^L \half A \Dp{t}\v r^2
+ \v w \cross A_\lambda \v d_\lambda \cdot \Dp{t}\v r
+ \half \v w \t I \v w
\hspace{1mm} ds
\\
\fancy{V}&=\int_0^L \half (\v u - \vh u) \t K (\v u -\vh u)
+ (\v u - \vh u) \t T (\v v - \vh v)
+ \half (\v v - \vh v) \t J (\v v - \vh v)
\hspace{1mm} ds
\\
\fancy{L}&=\fancy{T} - \fancy{V}
}
\end{equation}
where
$\Dp{s}\v r = \v u$,
$\Dp{s}\v d_i = \v u \cross \v d_i$,
$\Dp{t}\v d_i = \v w \cross \v d_i$,
$\t K=K_{ij} \v d_i \otimes \v d_j$,
$\t J=J_{ij} \v d_i \otimes \v d_j$,
$\t T=T_{ij} \v d_i \otimes \v d_j$ and
$\t I=I_{ij} \v d_i \otimes \v d_j$.
$t$ is time and $s$ is a label for the position of a
cross-section along the rod length (defined to run from $0$ to $L$
so that $\Delta s=1$ corresponds to one unit in the rest shape).
$\vh u$ and $\vh v$ define the shape of the rod
when no force is applied anywhere (reference configuration).
We use Einstein summation convention
where Greek letters are $1$ or $2$ and Roman letters range from $1$ to $3$
for the rest of this paper and repeated indices indicate sums eg.
$x_i y_i := \sum_{i=1}^{3} x_i y_i$.\footnote{
The tensor quantities above take the following values:
$K_{\lambda\gamma}=\delta_{\lambda\gamma} G,
K_{\lambda 3}=K_{3 \lambda}=0,
K_{33}=E$,
$J_{\lambda\gamma}=\delta_{\lambda\gamma} E I_{\lambda\gamma},
J_{\lambda 3}=J_{3 \lambda}=0,
J_{33}=G I_{33}$,
$T_{\lambda\gamma}=T_{33}=0,
T_{13}=-E A_2, T_{23}=E A_1, T_{31}=G A_2, T_{32}=-G A_1$
and $I_{\lambda\gamma}=\delta_{\lambda\gamma}A_{\mu\mu}-
A_{\lambda\gamma}, I_{\lambda 3}=I_{3\lambda}=0, I_{33}=A_{\mu\mu}$.
Here $A$ is the cross-sectional area, $A_{\lambda}$ its first
mass moments and $A_{\lambda\gamma}$ are its second mass moments.
}

Finding the stationary point of the Lagrangian through the usual
method with two variables yields the following coupled equations
(three dimensions each leading to a six dimensional, second order
linear PDE)
\begin{equation}
\label{eqn:Cosserat}
\eqalign{
\Dp{s}\vect{n} + \vect{f} = \rho A \Dp{tt} \vect{r} +
\Dp{tt}\vect{q}
\\
\Dp{s}\vect{m} + (\Dp{s}\vect{r}) \cross \vect{n} + \vect{l}
=\vect{q} \cross (\Dp{tt} \vect{r}) + \Dp{t} \vect{h}
}
\end{equation}
where $\v f$ and $\v l$ are external forces and torques and
$\vect{n}(s,t) = \Dp{\vect{v}} \fancy{V}(s,t)$,
$\vect{m}(s,t) = \Dp{\vect{u}} \fancy{V}(s,t)$,
$\vect{q}=A_{\gamma}\vect{d}_\gamma$ and
$\vect{h}=\t I \v w$.

As in \cite{Wang:Cosserat2D} we will adopt a quasi-static assumption.
This means that we assume that all time-dependent terms in
(\ref{eqn:Cosserat}) (i.e. the RHS) are set to zero but
the boundary conditions vary with time.
In the context of MEMS modelling this assumption can be justified
be realising that the frequencies of the entire, multi-component
system, are significantly lower than the vibrational frequency
of each component. This means that the wavelength of eigen modes
is greater than half the length of the rods.
With this assumption we reduce (\ref{eqn:Cosserat}) to a
six-dimensional, second order ODE which is much more amenable
to analytic work than the complete PDE.

\section{Material Perturbations}

In order to introduce a defect we can introduce changes to any (or all)
of: the shear and elastic moduli $G$ and $E$; the shape of the cross-section
changing $A$, $A_{\lambda}$ and $\t I$; and the reference configuration
through changes to $\vh u$ and $\vh v$.
Our technique involves introducing any of these changes as a
perturbation to the ideal case. That is, we find a closest possible
ideal case and treat differences as small. We can then expand
(to increasingly higher orders if we like) the Cosserat
equation in terms of the perturbation coefficient (which
we denote $\Gamma$) of the defect. The expansion in terms
of $\Gamma$ should then approach the complete solution as we
increase the number of terms.

Noticing that any changes to the material will only affect
$\v n$ and $\v m$, we may rewrite (\ref{eqn:Cosserat}) as
(to first-order in $\Gamma$)
$\v n(\v v, \v u)= \v n\ind{0} (\v v, \v u)
+ \Gamma \v n\ind{0}(\v v, \v u)$ and
$\v m(\v v, \v u)= \v m\ind{0} (\v v, \v u)
+ \Gamma \v m\ind{0}(\v v, \v u)$
and then solve the Cosserat equations as a perturbation
to the known system with $\v n=\v n\ind{0}$ and $\v m=\v m\ind{0}$.

Keeping only the first-order terms and maintaining the
quasi-static assumption we can now write
\begin{equation}
\label{eqn:CossPert}
\eqalign{
\Dp{s} \v n\ind{1} = 0
\\
\Dp{s} \v m\ind{1}
+ \t A(\v v\ind{0}) \cdotc \v n\ind{1}
+ \t A(\v v\ind{1}) \cdotc \v n\ind{0}
= 0
}
\end{equation}
where
\begin{equation}
\label{eqn:CossPertMN}
\fl
\eqalign{
\v n\ind{1} &=
\t K\ind{0} \cdotc \dv v\ind{1}
+ \t K\ind{1} \cdotc \dv v\ind{0}
+ \comm{\t A(\v x\ind{1})}{\t K\ind{0}} \cdotc \dv v\ind{0}
+ \dv u\ind{0} \cdotc \t T\ind{1}
\\
\v m\ind{1} &=
\t J\ind{0} \cdotc \dv u\ind{1}
+ \t J\ind{1} \cdotc \dv u\ind{0}
+ \comm{\t A(\v x\ind{1})}{\t J\ind{0}} \cdotc \dv u\ind{0}
+ \t T\ind{1} \cdotc \dv v\ind{0}
}
\end{equation}
and $\dv v\ind{I} = \v v\ind{I} - \vh{v}\ind{I}$,
$\dv u\ind{I} = \v u\ind{I} - {\vh u}\ind{I}$ and
$A(\v x\ind{1}) = \t R\ind{0} \t A(\vphi\ind{1}) \t R\ind{0}{}^T$
($\t R\ind{0}:=\t R(\phi\ind{0})=\v d_i\ind{0} \otimes \v e_i$).
Here the commutator term with $\t A(\v x\ind{1})$ takes into account
the effect of the change of basis set to first order.

This is then solved according to the boundary conditions
$\v r\ind{1}(0)= \v r\ind{1}(L)=0$ and
$\vphi\ind{1}(0)= \vphi\ind{1}(L)=0$.

\subsection{Series expansion in {$\epsilon$}}
As in \cite{Wang:Cosserat2D} let us make a
first-order expansion in $\epsilon$ (the
perturbation coefficient of the boundary conditions)
and substitute it into (\ref{eqn:CossPertMN}).
We can now write
\begin{eqnarray*}
\dv v\ind{0}=\epsilon \dv v\ind{0,1},
\dv u\ind{0}=\epsilon \dv u\ind{0,1} \\
\dv v\ind{1}=\dv v\ind{1,0} + \epsilon \dv v\ind{1,1},
\dv u\ind{1}=\dv u\ind{1,0} + \epsilon \dv u\ind{1,1}
\end{eqnarray*}
Here the first upper index refers to powers of $\Gamma$ and
the second to powers of $\epsilon$ and
$\v v\ind{0,0}={\vh v}$. We can justify the exclusion of terms
above $O(\epsilon)$ through the following argument:
the expression we use for the ideal case is accurate to
$O(\epsilon^3)$; our expression is accurate to $O(\Gamma \epsilon)$;
assuming manufacturing to be quite precise then
$\Gamma \aeq O(\epsilon^2)$; thus the total accuracy
for the non-ideal case $\aeq O(\epsilon^3)$

Under these assumptions, equations (\ref{eqn:CossPert}) and
(\ref{eqn:CossPertMN}) become
$\Dp{s} \v n\ind{1,0}=0$,
$\Dp{s} \v n\ind{1,1}=0$ and
\begin{eqnarray*}
\Dp{s} \v m\ind{1,0}
&+ \t A(\v v\ind{0,0}) \cdotc \v n\ind{1,0}
+ \t A(\v v\ind{1,0}) \cdotc \v n\ind{0,0}
= 0
\\
\Dp{s} \v m\ind{1,1}
&+ \t A(\v v\ind{0,1}) \cdotc \v n\ind{1,0}
+ \t A(\v v\ind{0,0}) \cdotc \v n\ind{1,1}
\\
& + \t A(\v v\ind{1,1}) \cdotc \v n\ind{0,0}
 + \t A(\v v\ind{1,0}) \cdotc \v n\ind{0,1}
= 0
\end{eqnarray*}
where
\begin{eqnarray}
\v n\ind{1,1} &=
\t K\ind{0,0} \cdotc \dv v\ind{1,1}
+ \t K\ind{1,0} \cdotc \dv v\ind{0,1}
+ \dv u\ind{0,1} \cdotc \t T\ind{1,0}
\nonumber\\ &
+ \comm{\t A(\v x\ind{1,0})}{\t K\ind{0,0}} \cdotc \dv v\ind{0,1}
+ \comm{\t A(\v x\ind{1,1})}{\t K\ind{0,0}} \cdotc \dv v\ind{0,0}
\label{eqn:PertPertN}\\
\v m\ind{1,1} &=
\t J\ind{0,0} \cdotc \dv u\ind{1,1}
+ \t J\ind{1,0} \cdotc \dv u\ind{0,1}
+ \t T\ind{1,0} \cdotc \dv v\ind{0,1}
\nonumber\\ &
+ \comm{\t A(\v x\ind{1,0})}{\t J\ind{0,0}} \cdotc \dv u\ind{0,1}
+ \comm{\t A(\v x\ind{1,1})}{\t J\ind{0,0}} \cdotc \dv u\ind{0,0}.
\label{eqn:PertPertM}
\end{eqnarray}

\section{Examples}

\subsection{`Blobs' and `Nicks'}
Sometimes a rod might be created with a bump of extra material or a
nick through errors in the manufacturing process. This would locally
affect the mass moments of the rod so that we would have to use
$A(s)$, $A_{\gamma}(s)$ and $A_{\gamma\lambda}(s)$ to calculate
$\t K$, $\t J$ and $\t T$.

Treating this defect as a small perturbation from an otherwise
uniform cross-section we can write
\[
\t K = \t K\ind{0,0} + \Gamma \t K\ind{1,0}(s), \spc
\t J = \t J\ind{0,0} + \Gamma \t J\ind{1,0}(s), \spc
\t T = \Gamma \t T\ind{1,0}(s)
\]
where $\t K\ind{0,0}=K\ind{0,n0}_{ij}
\v d_i\ind{0,0} \otimes \v d_j\ind{0,0}$ and
$\t K\ind{1,0}=K\ind{1,0}_{ij}(s) \v d_i\ind{0,0} \otimes \v d_j\ind{0,0}$
while $K\ind{1,0}_{ij}(s)=K_{ij}\ind{1,0}\N{s}$ (same for $\t J$ and $\t T$).
Note that $\t T\ind{0,0}=0$ as we choose the path of the ideal rod
to follow the centre-of-mass of the fixed cross-sections.

Let us find a solution to (\ref{eqn:CossPert}) where
the rod deviates slightly from its reference frame (to order
$\epsilon$) due to displacements and rotations of its end points.
We substitute
\begin{eqnarray*}
\dv v\ind{0,0} = \epsilon \v v\ind{0,1},
\dv u\ind{0,0} = \epsilon \v u\ind{0,1} \\
\dv v\ind{1,0} = \epsilon \v v\ind{1,1},
\dv u\ind{1,0} = \epsilon \v u\ind{1,1}
\end{eqnarray*}
into (\ref{eqn:PertPertN}) and (\ref{eqn:PertPertM})
(here the terms involving commutators vanishes as
$\t A(\v x\ind{1,0})$ is by
definition $O(\epsilon)$ and must therefore be
multiplied by $\dv v\ind{0,0}=0$) as follows
\begin{eqnarray*}
\v n\ind{1,1} &=
\t K\ind{0,0} \cdotc (\v v\ind{1,1} - \hv v\ind{1,1})
+ \t K\ind{1,0} \cdotc (\v v\ind{0,1}  - \hv v\ind{0,1})
+ \t T\ind{1,0}{}^\T \cdotc \v u\ind{0,1}
\\
\v m\ind{1,1} &=
\t J\ind{0,0} \cdotc \v u\ind{1,1}
+ \t J\ind{1,0} \cdotc \v u\ind{0,1}
+ \t T\ind{1,0} \cdotc (\v v\ind{0,1}  - \hv v\ind{0,1})
\end{eqnarray*}
which may then be used to calculate $\v v\ind{1,1}$ and
$\v u\ind{1,1}$.

After some work we can obtain expressions for
$\vphi\ind{1,1}$ and $\v r\ind{1,1}$ were we have made an assumption
that the shape of the perturbations distribution is governed
by the same function of $s$ for
each of $\t K$, $\t J$ and $\t T$ (this is completely true for
nicks and blobs). The solutions are thus given by
\begin{equation}
\eqalign{
\fl \t J\ind{0,0} \vphi\ind{1,1} =
\v k_m\ind{1,0}s - \t A_3 \cdotc \v k_n\ind{1,0} \frac{s^2}{2}
\\
- \t J^{\prime}
\lbr \v k_m\ind{0,0} \tilde{1}
- \t A_3 \cdotc \v k_n\ind{0,0} \tilde{s} \rbr
- \t T^{\prime} [\t K\ind{0,0}]^{-1} \v k_n\ind{0,0} \tilde{1}
\\
\fl \t K\ind{0,0} \v r\ind{1,1} =
\t K\ind{0,0} \t A_3 \cdotc \int_0^s \v  \phi\ind{1,1} d s^{\prime}
+
\v k_n\ind{1,0} s - \t K^{\prime} \cdotc \v k_n\ind{0,0} \tilde{1}
\\
- \t T^{\prime}{}^\T [\t J\ind{0,0}]^{-1} \cdotc
(\v k_m\ind{0,0}\tilde{1} - \t A_3 \v k_n\ind{0,0}\tilde{s})
}
\end{equation}
where $\tilde{f}(s)=\int_0^s f(s^{\prime}) \N{s^{\prime}} ds^{\prime}$.
We can calculate $\v k_n\ind{1,0}$ and $\v k_m\ind{1,0}$
($\v k_n\ind{0,0}$ and $\v k_m\ind{0,0}$ are known from the ideal case)
by ensuring that $\vphi\ind{1,1}(L)=0$ and $\v r\ind{1,1}(L)=0$.
The full analytic solutions are far too complicated for inclusion here
although the MAPLE source to generate them is available on request.

\begin{figure}
\includegraphics[scale=0.37]{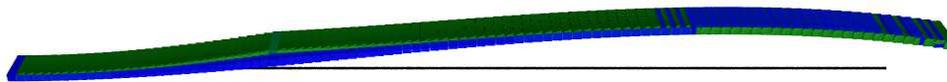}\\
\caption{Model of a beam demonstrating its shape
when it contains/does not contain a defect. The blue represents the
ideal case while the green beam has a defect located at the marked
point.}
\label{fig:BlobBeam}\
\end{figure}

Figure \ref{fig:BlobBeam} shows a visual demonstration of the model
were we have calculated the bending properties of a beam of certain
dimensions and boundary conditions both with and without a defect.
The difference is, as expected, greatest around the position of the
defect itself. Both the change in shape, and potential energy
from the defect itself will affect the restoring force of the beam.

\subsection{Random jitter}

The treatment of random jitter is all but identical to that
of a localised defect such as a blob or nick. The only difference
is that the distribution function $\N{s}$ will have different
expectation values which depend on the type and magnitude of the noise
(eg. we may have $\overline{1}=\int_0^L \N{s} 1 ds=0$
whereas $\overline{s}=k$). Substituting these values into
the final expressions (which we have not included here) we can
calculate the behaviour of the rod. Although beyond the scope
of this paper, the same model will also be applicable to the
calculation of noise `distributions'.

This second example demonstrates the power of the
technique as the same equations can be used for two different
types of physical impurities. Other impurities can be
modelled in similar ways by changing the `noisy' variable.

\section{Implementation}

One of the main advantages of the method outlined in \cite{Wang:Cosserat3D}
is that it gives us a means to convert the internal structure
of a rod into a generalised force dependent only on the
value of the end points (position and orientation of each end).
This allows us to easily generate a component model of the
rod for use in a component level simulator.

Using MAPLE we can follow the same
procedure to derive analytic expressions for the
change to the effective spring matrix of the end points
(a $12 \times 12$ matrix) caused by a defect.
We then add our new term to the third-order expression
from \cite{Wang:Cosserat3D} and create a VHDL-AMS file
representing the defective rod.
This is then compiled through a component level simulator
(SMASH 5.2.0\tm \footnote{SMASH is trademarked to Dolphin Integration})
and can be connected to other components.

The expression for the change in potential energy from the
defect can be written as
\begin{eqnarray}
\fl
\fancy{V}\ind{1,0}=\int_0^L [
\v n\ind{1,0} \cdot \dv v\ind{0,0}
\v m\ind{1,0} \cdot \dv u\ind{0,0}
- \half \dv v\ind{0,0} \t K\ind{1,0} \dv v\ind{0,0}
\nonumber \\
\lo
- \half \dv u\ind{0,0} \t J\ind{1,0} \dv u\ind{0,0}
- \half \dv u\ind{0,0} \t T\ind{1,0} \dv v\ind{0,0}
] ds
\end{eqnarray}
and we ignore the change to kinetic energy so that
$\v F\ind{1,0} = \t K\ind{1,0} \v Q$ where
$\t K\ind{1,0} = \Dp{\v Q}\Dp{\v Q} \fancy{V}\ind{1,0}$.
Here $\v Q:=[x_1, y_1, z_1, \phi_{x1}, \phi_{y1}, \phi_{z1},
x_2, y_2, z_2, \phi_{x2}, \phi_{y2}, \phi_{z2}]$ where $x_1$ etc.
are the changes to the boundary conditions at the two ends
of the beam.

\section{Results}

\begin{figure}
{
\tiny
\begin{tabular}{p{60mm}p{60mm}}
\includegraphics[scale=0.55]{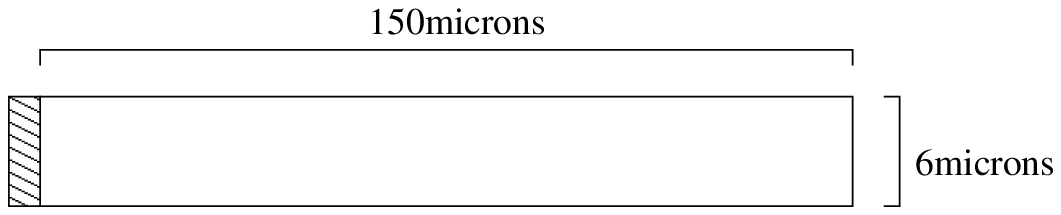}
\parbox{40mm}{Case I - Ideal case with one end fixed.}
&
\includegraphics[scale=0.55]{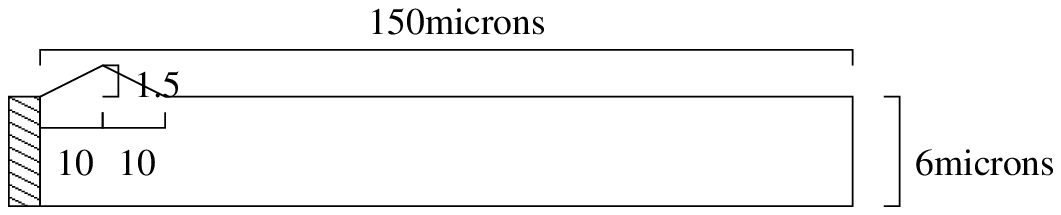}
\parbox{40mm}{Case II - As Case I with a blob near the fixed end.}
\\
\includegraphics[scale=0.55]{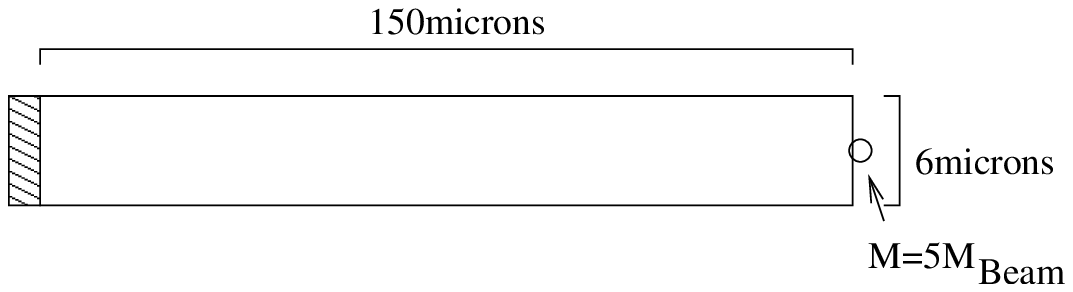}
\parbox{40mm}{Case III - Ideal case where a mass of $0.1573\rm{ng}$ is
attached to the free end.}
&
\includegraphics[scale=0.55]{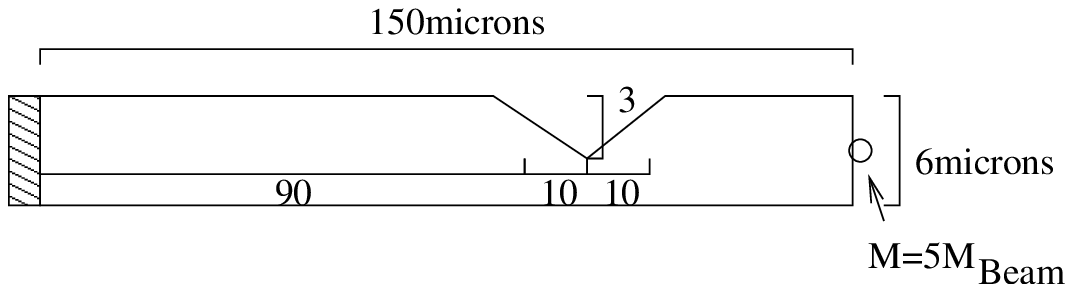}
\parbox{40mm}{Case IV - As Case II with a nick of depth $3$ at $100$\micron.}
\\
\includegraphics[scale=0.55]{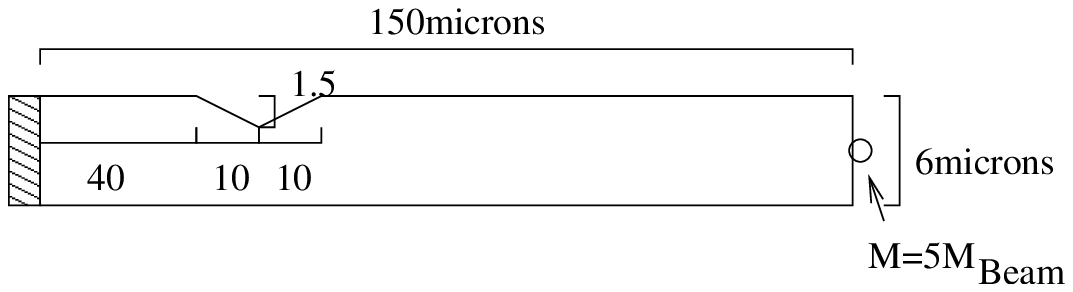}
\parbox{40mm}{Case V - As Case II with a nick of depth $1.5$ at $50$\micron.}
&
\includegraphics[scale=0.55]{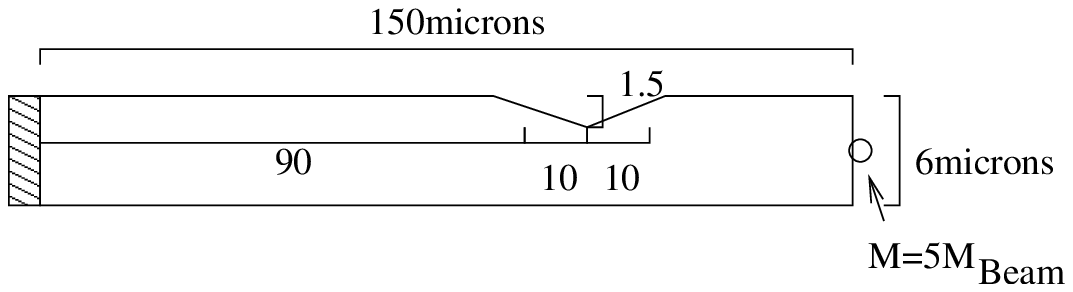}
\parbox{40mm}{Case VI - As Case III with a nick of depth $1.5$.}
\end{tabular}
}
\caption{Diagrams of the Test Cases. Each of these is a
variant of Case I and the height ($15$\micron, not pictured) remains
unchanged throughout. In all cases we consider the lowest energy
mode where the motion is in the plane of the diagrams.}
\label{Fig:Beams}
\end{figure}

In order to verify the applicability and accuracy of the defect model
(and for that matter the ideal Cosserat model) we compare the results
of FEA simulations with those of the component model. We compare
six different slender beams, each a variation of a
length $150$\micron, width $6$\micron\ and height $15$\micron\ beam
(see Figure \ref{Fig:Beams}). These dimensions are quite typical of
MEMS devices although we use a shorter beam to ensure that the
FEA calculations run quickly (about one hour on a
2.80GHz Pentium 4 with 1GB RAM for the longest calculations)
with a decent accuracy. All simulations are performed in
FELT \cite{Felt} and it is worth noting that the solution of
the Cosserat model takes an imperceptible amount of time on
the same computer.

\begin{figure}
\begin{center}
\includegraphics[scale=0.5]{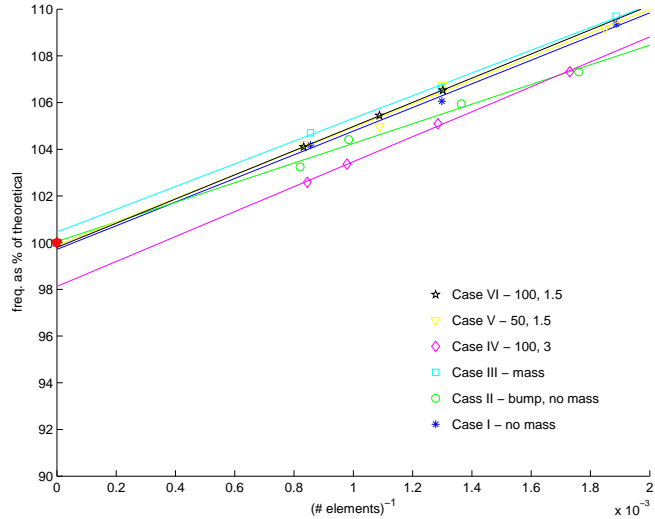}
\end{center}
\caption{Graphs of FEA calculations and their extrapolated values
where the frequencies are renormalised to a percentage of
the Cosserat value.}
\label{Fig:Fits}
\end{figure}

As a first test we must determine the accuracy of the Cosserat model of
an ideal beam. For this purpose we must consider the result of a
`perfect' (i.e. one with an infinite number of elements in the beam)
FEA calculation. For this purpose we must extrapolate the results
of FEA calculations to infinity. We assume that for any FEA calculation,
the result will take the form
$f_{\rm{FEA}}=f_{\infty} + \sfrac{\delta}{\textrm{(number of triangular elements)}}$.
With this assumption it is possible to calculate $f_{\infty}$ by extrapolating
a number of calculations with varying numbers of elements. This
involves making a linear fit to $f_{\rm{FEA}}$ versus
$\sfrac{1}{\rm{\# elements}}$. Figure \ref{Fig:Fits} demonstrates
the validity of this method. It is quite apparent from the graphs
that the FEA results lie close to a straight line for all six cases.
Of further assurance is the fact that $\delta$ is positive in
all cases. This is as expected since an FEA calculation will
necessarily be more restrictive than the infinite case and must
produce a higher lowest-mode frequency.

We summarise the results of our calculations in Table \ref{Tab:Results}.
Considering first the two ideal cases (I and III, without and with an
attached mass respectively) we see that the Cosserat model is a highly accurate model of a beam. In both cases it agrees to within
$0.5$\% of the extrapolated FEA result ($f_{\infty}$).
This is well within the error bounds of the extrapolation technique.

\begin{table}[h]
\begin{tabular}{p{10mm}rrr}
System &
\multicolumn{1}{l}{$f_{\infty}$} &
\multicolumn{1}{l}{$f_{\rm{Coss}}$}
& \% Err
\\
& (GHz) & (GHz) \\
\hline
I & 334.2 & 335.3 & 0.29
\\
II & 358.3 & 358.1 & -0.045
\\
III & 72.18 & 71.85 & -0.46
\\
IV & 69.16 & 70.48 & 1.9
\\
V & 69.28 & 69.37 & 0.13
\\
VI & 71.08 & 71.22 & 0.21
\end{tabular}
\caption{Results of the FEA and Cosserat Simulations.}
\label{Tab:Results}
\end{table}

Moving on to the defect model we observe that the results of the
Cosserat model or similarly close to those of the FEA simulations
for all cases but IV. The `poor' accuracy of Case IV is easily
understood by considering that we have chosen an extreme case
which goes beyond the expected range of the perturbation
technique used. A defect this large (half the width of the beam)
would not occur in a real MEMS device and would almost certainly
cause significant problems (such as a full breakage)
beyond the applicability of any beam model.

\section{Conclusion}

In this paper we have developed a technique for the construction of
beam models with defects. We apply this technique to the construction
of a model of a beam with a blob/nick present as well as
a model of a beam with random jitter.

The theoretical framework presented can be effectively
applied to new cases for the development of new models.
This allows us to devise models for defects which are
presently unobserved (such as kinks in a beam), but
which may cause problems in future devices, particularly those
made of novel materials or with novel applications.

The developed blob/nick and random jitter models could be used for
reliability tests of MEMS both at the design and production stage.
This is particularly valuable in the modelling of high
specification MEMS, where an accurate model of faults is vital
for the design process as costs are high for experiments.

Tests on the blob/nick model show that the model is a highly
accurate tool for the simulation of such systems. An accuracy of
at worst $0.5$\% for all but the most extreme case (which would
not appear in practice) suggests that the model is more than
satisfactory for simulation of real devices through
Component Level Simulations or FEA simulations.

One possible future use of this technique would be to develop a model of
a curved beam with a defect. At present, curved beams present
difficulties for standard techniques of ideal beams. The
Cosserat model has already been applied to curved beam
models \cite{Grav} and application of the aforementioned
technique to this work may allow accurate modelling of future
MEMS devices such as accelerometers.

\section*{Acknowledgements}
We would like to thank Richard Rosing at Lancaster University and
colleagues at QinetiQ (Malvern) and ST Microelectronic (Milan)
for
helpful discussions. The work is funded by the EPSRC under the Computational
Engineering
Mathematics Programme.

\bibliographystyle{unsrt}

\bibliography{paper}

\end{document}